# Detecting Coordination Problems in Collaborative Software Development Environments

Chintan Amrit & Jos van Hillegersberg

University of Twente, The Netherlands

{c.amrit, j.vanHillegersberg}@utwente.nl

**Abstract**

*Software development is rarely an individual effort and generally involves teams of developers collaborating to generate good reliable code. Among the software code there exist technical dependencies that arise from software components using services from other components. The different ways of assigning the design, development, and testing of these software modules to people can cause various coordination problems among them. We claim that the collaboration of the developers, designers and testers must be related to and governed by the technical task structure. These collaboration practices are handled in what we call Socio-Technical Patterns.*

*The TESNA project (Technical Social Network Analysis) we report on in this paper addresses this issue. We propose a method and a tool that a project manager can use in order to detect the socio-technical coordination problems. We test the method and tool in a case study of a small and innovative software product company.*

# 1. Introduction

The classic paper by Conway (Conway, 1968) states organizations which design systems are constrained to produce designs which are copies of the communication structure of these organizations. In other words, if the teams involved in software production have shortcomings in their interpersonal relationships, the resulting technical architecture of the software is likely to be flawed. Since Conway, researchers have identified various more detailed patterns that describe the preferred relationships between team communication structure (the social network) and technical software architecture. We define such patterns as socio-technical patterns. Socio-Technical patterns can be used in addition to the widely known Technical Patterns, that guide the system design and architecture (Gamma, Helm, Johnson, & Vlissides, 1995), and Social Patterns that guide optimal team structure and organization (Yang & Tang, 2004). Currently a growing collection of Socio-Technical Patterns exists suggesting optimal team compositions and task assignments depending on the modularity of the technical architecture (Cockburn, 1996; J. Coplien, O. & Harrison, 2004). However, these patterns have usually not been empirically validated and are hard to implement and monitor in practice. Especially within dynamic globally distributed environments, the use of socio-technical patterns is challenging.

The TESNA project (Technical Social Network Analysis) we report on in this paper addresses this issue. We have developed a method and a tool that a project manager can use in order to detect Socio-Technical Structure Clashes (STSCs). An STSC occurs if and when a Socio-Technical Pattern exists that indicates that the social network of the software development team does not match the technical dependencies within the software architecture under development. We claim that continuous and early detection of STSCs can help project managers in monitoring the software development process and enable them to take actions whenever a STSC occurs. We test the method and tool in a case study of a small and innovative software product company. Our approach is illustrated in Figure 5. We explain the idea of STSC in more detail below and illustrate how the TESNA method and tool support the project manager to implement Socio-Technical Patterns and use them to redesign the social network and/or the technical architecture. The case study also provides us with feedback on the applicability of various Socio-Technical Patterns in the context of a dynamic software product development environment.

The rest of the paper is structured as follows; section 2 provides a background of Patterns and Socio-Technical Patterns, section 3 provides a literature overview, section 4 describes the research site and the methods that we have used to conduct our research, section 5 describes the Conway's law pattern and it validation with the help of the CTO's feedback, section 6 explains the betweenness centrality pattern and finally section 7 the conclusion and future work.

## 2. Patterns Background

Software Development projects often prove to be both a costly and risky endeavour. Poor software project execution continues to result, in the best cases, in missed deadlines, and in the worst cases, in escalations in commitment of additional resources as a cure-all for runaway projects (Kraut & Streeter, 1995).

Some of these problems stem from the differences between the process model and software architecture at the project planning phase to what actually occurs at the development phase(Curtis, Krasner, & Iscoe, 1988). Curtis et al (Curtis et al., 1988) describe the project manager's predicament when there are changes in the software application and related technologies due to fluctuating specifications or requirements. They describe how the tracking schemes most managers had developed were of no use and they had to rely on system engineers for their managerial input.

Fig 1 describes the overview of this problem. On the left hand side of Fig 1 we have the design phase where the software process model as well as the software architecture are planned and designed while on the right hand side the actual implementation of the software development is described. The implemented software evolves in to something completely different from what was envisioned at the design phase over a period of time, as shown in Fig 2. In order to develop and maintain quality software in a repeatable predictable fashion and to prevent the software process from getting out of hand the industry has what are called software best practices. These best practices are commercially proven approaches to strike at the root of the software development problems (Kruchten, 1998). During software development, just knowledge of the best practices is not enough to guarantee successful completion of software projects. The problem with the usage of best practices as generic solutions is that they are not precisely formalized and hence not easily applicable. What are needed are generic solutions to specific problems one encounters during the actual practice of software development. Experienced software designers and developers try to reuse solutions that have worked in the past rather than solve every problem from first principles. This methodology has led to the use of software patterns, which are proven solutions to recurrent software development problems. These patterns are applied in the design stage of the product (Fig 1) and are not applied in the actual implementation part of the software development.

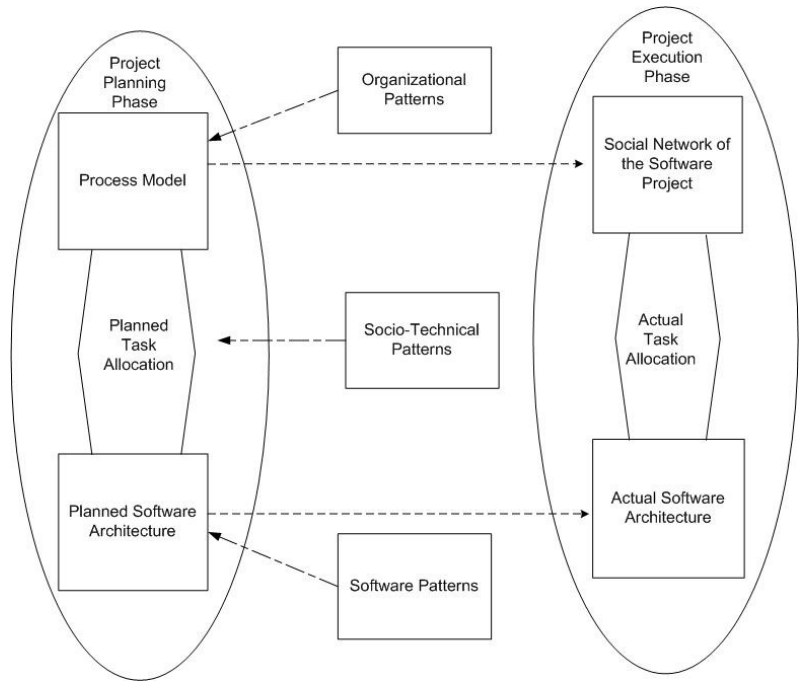

Figure 1: Pattern usage during the Project Planning

## 2.1 Different Patterns in Software Development

While there are many ways to describe patterns, Christopher Alexander who originated the notion of patterns in the field of architecture described patterns as a recurring solution to a common problem in a given context and system of forces (Alexander, Ishikawa, & Silverstein, 1977). In Software Engineering patterns are attempts to describe successful solutions to common software problems (Schmidt, Fayad, & Johnson, 1996). Patterns reflect common conceptual structures of these solutions and can be used repeatedly when analyzing, designing and producing applications in a particular context.

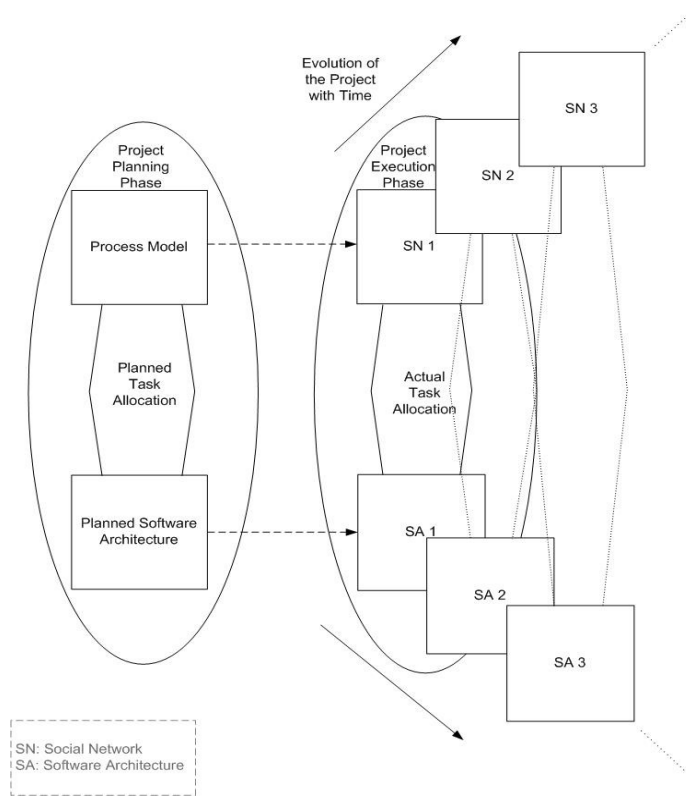

Figure 2: The evolution of the project with time

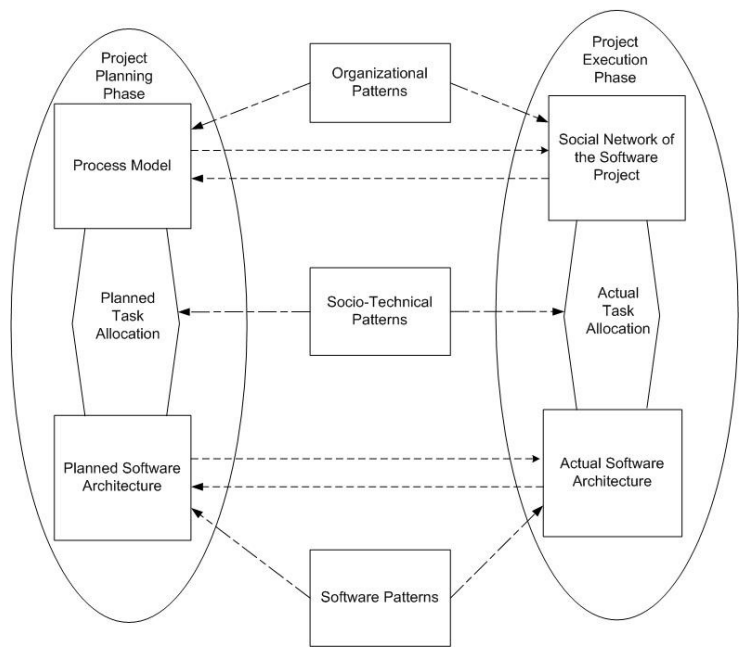

Figure 3: Pattern usage at both the Planned and Execution stages can help Project Management

Patterns represent the knowledge and experience that underlie many redesign and re-engineering efforts of developers who have struggled to achieve greater reuse and flexibility of their software. The different types of patterns are:

- Design Patterns: Are simple and elegant solutions to specific problems in software design (Gamma et al., 1995).
- Analysis Patterns: Capture conceptual models in an application domain in order to allow reuse across applications (Fowler, 1997).
- Organizational Patterns: Describe the structure and practices of human organizations (J. Coplien, O. & Harrison, 2004).
- Process Patterns: Describe the Software Design Process (J. Coplien, O. & Harrison, 2004)

The basic format of a pattern was devised by the "Gang of Four" (Gamma et al., 1995), and can be seen on the first column in Table 1.

| **Pattern Name** | **Conway's Law (Conway, 1968)** | **Betweenness centrality match (Hossain, Wu, & Chung, 2006)** |
|---|---|---|
| **Problem:** A problem growing from the Forces | Aligning Organization and Architecture | Centrality of important people |
| **Context:** The current structure of the system giving the context of the problem | An architect and development team are in place | Social Network of the team at different stages of software development |
| **Forces:** Forces that require | Architecture shapes | People who are not central to |
| resolution | communication paths in the organization. Formal Organization shapes Architecture | the software development or management take a central role in coordination |
| **Solution:** The solution proposed for the problem | Make sure organization is compatible with the architecture | Make sure the important people take a more central role in coordination. |
| **Resulting Context**: Discusses the context resulting from applying the pattern. In particular, trade-offs should be mentioned | The organization and product architecture will be aligned. | Project critical information will be conveyed to all team members. |
| **Design Rationale/Related patterns:** The design rationale behind the proposed solution. Patterns are often coupled or composed with other patterns, leading to the concept of pattern language. | Historical | Betweenness centrality is a key predictor for coordination |

Table 1: The Socio-Technical Patterns used in this paper

**2.2 Socio-Technical Patterns**

Some of the problems concerning development activities have been collected and described by Coplien et al (J. Coplien, O. & Harrison, 2004) including a set of what they call Process

Patterns to deal with these coordination problems. As the term process patterns is also used in business process management and workflow, we prefer to use the term *Socio-Technical Patterns* to refer to those patterns involving problems related to both the social and technical aspects of the software process. As they capture a wide variety of knowledge and experience, Socio-Technical Patterns are potentially very useful to aid the project manager in planning and monitoring a complex development project. However, these patterns have not been extensively validated empirically and can be hard to implement. The reason why the patterns are difficult to implement is that the problems addressed by the patterns are hard to detect, as existing techniques are labour intensive. The lack of empirical validation makes it difficult for the project manager to decide on which Socio-Technical pattern to apply to his project.

## 3. Literature Overview and Research Focus

There are three kinds of structure clashes, those at the social level (where the planned process model doesn't match the actual social network Fig 1.), those at the technical level (where the actual software architecture doesn't match the planned Fig 1.) and those at the socio-technical level (where the planned task allocations don't match the actual Fig 1.). Table 2 provides an overview of some of the important literature in each field. As the amount of literature on structure clashes in the social and in the technical levels is a lot, the table is not exhaustive.

In this paper we focus on Socio-Technical clashes. Though these clashes are present as patterns in literature (J. Coplien, O, 1994; J. Coplien, O. & Harrison, 2004), these patterns are not always applied in the implementation phase of software development. Over a period of time the designed process model evolves into a social network of developers with a different task allocation than that planned at the design phase. The software architecture also evolves with time and becomes very different from what was envisioned at the design phase (Guo, Atlee, & Kazman, 1999; G. Murphy, C. , Notkin, & Sullivan, 2001) (Fig 2). This is a problematic scenario, as the manager responsible has no control over the project anymore. This lack of control could lead to extensions and project overruns. In the case of Structure Clashes in the software architecture, one choice is to ignore transformation and to proceed with the task based on information from the source code. In the case of the gap in the process model one can continue with the development based on tasks assigned locally among the project teams. Though these strategies may work in small systems and teams, in larger development projects this could lead to inappropriate choices and delays in development (G. Murphy, C. et al., 2001) resulting in financial losses for the project.

Though there have been research works highlighting the gap between design and implementation in software architecture (Guo et al., 1999; G. Murphy, C. et al., 2001), there is not much research conducted in identifying and remedying the gap between design and implementation in the organization and process of implementation of Software Development. While applying Software Patterns can keep the software architecture under managerial control (Guo et al., 1999), the same can also be done by applying Organizational and Socio-Technical patterns to the process and the planned task allocation (Fig 3). In this research we use Socio-Technical patterns in order to spot STSCs. Regular detection of these STSCs can help the manager apply Socio-Technical patterns to the software process model and thereby keep the software process evolution under control (Fig. 3). We approach the literature review of clashes in three separate sections, the purely technical, purely social and the socio-technical.

**3.1 Technical Structure Clashes**

The technical architecture of the software system may drift from the documented architecture if architecture changes are made during software implementation and no effort is made to maintain the architecture documents.

In the past, reverse engineering methods have been used to prevent the software architecture from drifting. One of the reverse engineering methods has been to extract the software's call-graph and compare it with the expected call-graph (Woods and Qiang 1995; Murphy, Notkin et al. 2001). Guo et al (Guo et al., 1999) describe a semi-automatic analyses that codifies heuristics (in accordance to Design Patterns) in order to apply existing reverse-engineering tools. Also, there is a number of reverse engineering tools developed to automatically extract, manipulate and query source model information. For example, Rigi (Tilley, Wong, Storey, & Muller, 1994), LSME (G. Murphy, C. & Notkin, 1996), IAPR (Kazman, 1998), RMTool (G. Murphy, C. et al., 2001) and Deli (Kazman & Carriere, 1998) are some of the reverse engineering tools used in practice. Though there are many tools for reverse engineering the software architecture we find very few tools to do that same with the Software Process Model.

| Papers | Technical Clashes | Social Clashes | Socio-Technical Clashes in Engineering | Socio-Technical Clashes in Software Engineering |
|---|---|---|---|---|
| Murphy, Notkin & Sullivan, 2001 | √ | | | |
| Guo, Yanbing & Atlee 1999 | √ | | | |
| Woods & Qiang 1995 | √ | | | |
| Faraj & Sproull, 2000 | | √ | | |
| Stewart & Barrick, 2000 | | √ | | |

| Reference | | | | |
|---|---|---|---|---|
| Yang &Tang, 2000 | | √ | | |
| Baldwin, Bedell & Johnson, 1999 | √ | | | |
| Sparrow, Liden, Wayne & Kramer, 2001 | | √ | | |
| Morelli, Eppinger, IEEE TEM1995 | | | √ | |
| Sosa & Eppinger 2004 | | | √ | |
| Wagstrom & Herbsleb, 2006 | | | | √ |
| Cataldo, Wagstrom, Herbsleb, 2006 | | | | √ |
| Ovaska, Rossi & Marttiin 2003 | | | | √ |
| MacCormack & Rusnack, 2004 | √ | | | √ |

Table 2: A Brief Overview of Important Literature

## 3.2 Social Network based Structure Clashes

Teams are the basic building block for many contemporary business organizations. Structure clashes are dealt with in organizational literature by focussing on how one can improve coordination in software development projects using the concepts of coordination between and among teams keeping task assignment as a moderating variable. Coordination refers to team-situated interactions aimed at managing resources and expertise dependencies (Faraj & Sproull, 2000). Research on software development teams has found that team performance is linked with the effectiveness of teamwork coordination (Kraut & Streeter, 1995).

Faraj and Sproull (Faraj & Sproull, 2000) take two perspectives on coordination: administrative coordination and expertise coordination. They claim that administrative coordination (management of tangible and economic resource dependencies) is good for simple routine tasks, while for complex non-routine intellectual tasks, expertise coordination (the management of knowledge and skill dependencies) become more important. Through expertise coordination the team can recognize and access expertise when it's needed.

Stewart and Barrick (Stewart & Barrick, 2000) build on organization-level findings and show that differences in how responsibilities are apportioned and coordinated correspond to variance in performance at the team level. They also show that the effect of these social elements is moderated by technical demands (tasks), consistent with socio-technical systems theory.

Sparrowe et al. (Sparrowe, Liden, Wayne, & Kraimer, 2001) hypothesize that centrality in a work group's advice network will be positively related to an individual's job performance. Where centrality in the advice network reflects an individual's involvement in exchanging assistance with co-workers and engaging in mutual problem solving. An individual who is central in the advice network is, over time, able to accumulate knowledge about task-related problems and workable solutions (Baldwin, Bedell, & Johnson, 1997). While the central individual develops problem solving capability and serves as a valued resource for future exchanges with co-workers, those individuals who are in peripheral positions in the advice network find it difficult to develop expertise and competencies for high levels of performance (Sparrowe et al., 2001). Hence, Sparrowe et al. (2001) hypothesize that centralization in a work group's advice network is negatively related to group performance.

Yang and Tang (Yang & Tang, 2004) try to analyse the relation between team structure and the performance of information systems development using a social network approach. They show how the structural properties of the work groups fluctuate during the various phases of software development, and how group cohesion and centrality are related to the final ISD performance. Though Yang and Tang (2004) do show how social research methods can be used to tackle "group process" factors, they do not deal with task allocation nor do they illustrate how one can solve the problem of task allocation among team members.

### 3.3 Socio-Technical Structure Clashes

An STSC (as described earlier) occurs if and when a Socio-Technical Pattern exists that indicates that the social network of the software development team does not match the technical dependencies within the software architecture under development. STSCs are thus indicative of coordination problems in a software development organization. We find a lot of literature in the organizational, production-engineering domain that deal with task allocation and coordination among the workers. While the use of Design Structure Matrices (DSM) to locate coordination problems in the field of software engineering is relatively less.

DSM (also known sometimes as Dependency matrices) have been used in Engineering literature to represent the dependency between people and tasks (Steven, Daniel, Robert, & David, 1994). Recent empirical work uses DSMs to provide critical insights into the relationship between product architecture and organizational structure. Morelli et al. (Morelli, Eppinger, & Gulati, 1995) describe a methodology to predict and measure coordination-type of communication within a product development organization. They compare predicted and actual communications in order to learn to what extent an organizations communication patterns can be anticipated.

Sosa et al (Sosa, Eppinger, & Rowles, 2004) find a "strong tendency for design interactions and team interactions to be aligned," and show instances of misalignment are more likely to occur across organizational and system boundaries. Sullivan et al. (Sullivan, Griswold, Cai, & Hallen, 2001) use DSMs to formally model (and value) the concept of information hiding, the principle proposed by Parnas to divide designs into modules (Parnas, 1972).

In the field of software engineering the application of DSM principles has been less and infrequent compared to other engineering domains. The following paragraphs give an overview of the literature in software engineering that deals with problems of coordination between people and technical tasks using DSM concepts.

DeSouza et al (de Souza, Redmiles, Cheng, Millen, & Patterson, 2004) describe the role played by APIs (Application Program Interfaces) which limit collaboration between software developers at the recomposition stage (R. Grinter, E., 1998).

Cataldo et al (Cataldo, Wagstrom, Herbsleb, & Carley, 2006) as well as Wagstrom and Herbsleb (Wagstrom & Herbsleb, 2006) do the same study of predicted versus actual coordination in a study of a software development project in a large company project. Their work provides insights about the patterns of communication and coordination among individuals working on tasks with dynamic sets of interdependencies.

Ovaska, Rossi and Marttiin (Ovaska, Rossi, & Marttiin, 2003) describe the role of software architecture in the coordination of multi-site software development through a case study. They suggest that in multi-site software development it's not enough to coordinate activities, but in order to achieve a common goal, it is important to coordinate interdependencies between the activities. The interdependencies between various components are described by the software architecture. So, if the coordination is done using the software architecture, the work allocation is made according to this component structure.

In splitting work along the lines of product structure one must consider the modular design of the product in order to isolate the effect of changes (Parnas, 1972). MacCormack and colleagues (MacCormack, Rusnak, & Baldwin, 2006) reiterate Conway's argument (Conway, 1968) when they compare commercial and open source development . As the software developers, in their study were collocated in the commercial project, it was easier to build tight connections between the software components, therefore producing a system more coupled compared to the similar open source project with distributed developers.

While the Conway's Law relation between the task and coordination of the developers has been validated by several empirical studies (Curtis et al., 1988; R. Grinter, E. , Herbsleb, & Perry, 1999; Herbsleb & Grinter, 1999; Sosa et al., 2004), we use this Conway's law as a means to identify a possible Structure Clash in the software development process.

**3.4 Identifying Socio-Technical Structure Clashes**

While there are many tools available for dealing with Technical Structure Clashes, there are few tools available for Socio-Technical Structure Clashes. Augur is a visualization tool that supports distributed software development process by creating visual representations of both the software artefacts and the software development activities (Froehlich & Dourish, 2004). de Souza et al. (de Souza et al., 2004) are developing a tool that checks dependency relationships between software call graphs and developers. Also there is a tool under development for forecasting dependencies between developers in an agile environment (Cataldo et al., 2006). These tools check for only one particular STSC and don't provide extensive software process re-engineering guidance.

Identifying the STSCs related to Socio-Technical patterns (J. Coplien, O, 1994; J. Coplien, O. & Harrison, 2004) can prove difficult for large distributed or collocated teams working on large software projects. These Socio-Technical patterns apply to ambitious, complex endeavours, that may comprise hundreds of thousands or millions of lines of code, while the size of the organizations considered range from a handful to a few dozen (J. O. Coplien & Schmidt, 1995). The Socio-Technical patterns can however be applied to larger organizations, if they are broken into smaller decoupled parts, where the patterns can be applied to the smaller parts of the organization. We contend that this problem related to a lack of control of the software project can be solved by a periodic assessment of STSCs.

In this paper we chose two particular Socio-Technical Patterns namely (*Conway's Law* and *Betweenness Centrality Match*) because we saw that these patterns were relevant for the case study in hand. We could have selected other Socio-Technical Patterns, for example, if we had selected *Code Ownership Pattern* (Pattern 15 from (J. Coplien, O, 1994)), the STSC we would have had to deal with would have been related to the problem this particular pattern addresses, namely, "*A developer cannot keep up with a changing base of implementation code*" (J. Coplien, O, 1994). The more concrete form of the STSC (according to the solution the pattern provides) is that: "*each code module must be owned by a single developer*". So the STSC we can check for is: all instances of code modules modified by multiple developers (with no single developer taking the responsibility) as well as code modules not modified by any developer for a very long period of time (especially if the code is not stable).

## 4. Conceptual Model

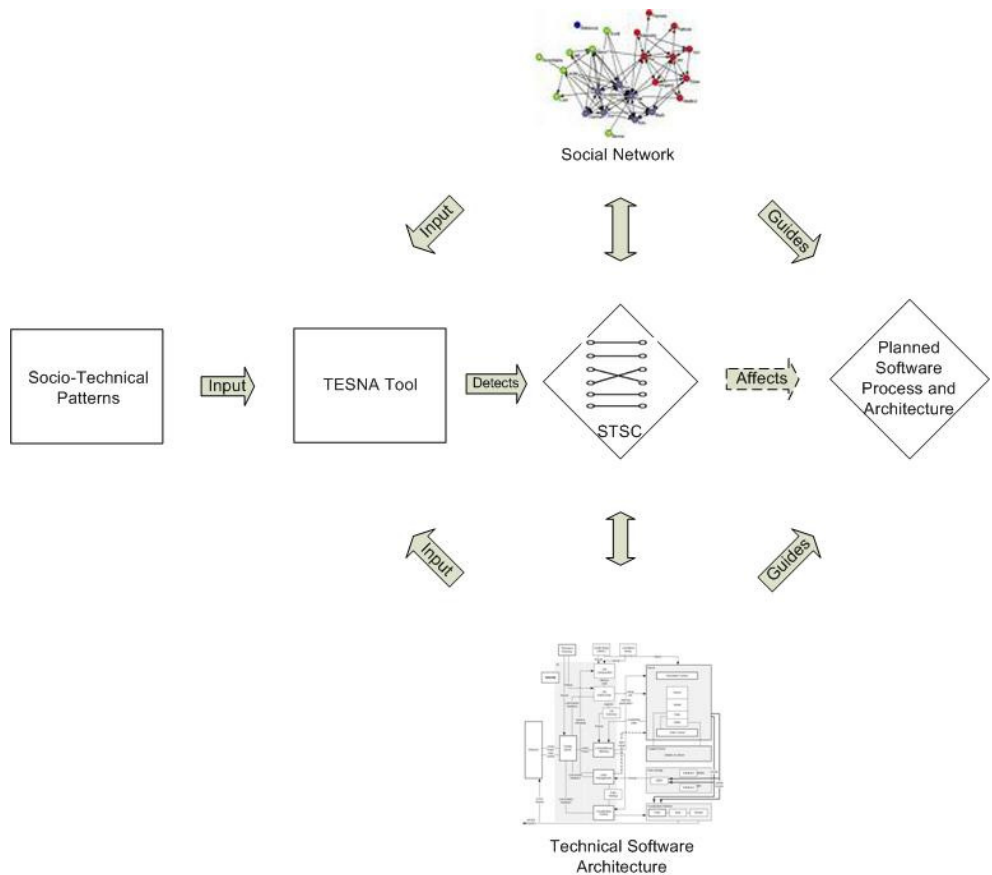

Figure 4: Socio-Technical Structure Clashes and the planned Software Process

Figure 4 represents the focus of this paper. We focus on the pattern implementation problem first. Our motivation is that when implementation and monitoring of patterns is enhanced, empirical validation of patterns will also become feasible. We provide a Method and Tool called TESNA (TEchnical and Social Network Analysis) that can improve the system development by regularly monitoring the software development project and alerting in case of a STSC. The Method consists of several steps. First, input from the Social Network as well as the Software Architecture is taken, and the tool TESNA provides a visual description of the networks and metrics, based on the Socio-Technical Patterns selected. The next step of the method lies in the detection of STSCs with the help of the visualizations that can help the manager to modify the informal design process model in order to improve project planning (Fig. 1). Currently the tool supports both qualitative as well as quantitative detection of STSCs, but for the purpose of this case study we used qualitative STSC detection methods. Furthermore, we selected two Socio-Technical Patterns: Conway's Law (Conway, 1968), and Betweenness Centrality Match (Hossain et al., 2006). Table 1 describes the two Socio-Technical Patterns used in this paper in greater detail. Please note that the STSC corresponds to the *Problem* each of the patterns address (according to the solution provided).

## 5. Research Site and Methods

Our case study was conducted in a software company called MENDIX, who are developing a large middleware product. One version of the middleware product was already released at the time of the study.

The reason behind choosing this case was that MENDIX followed an iterative incremental software development by incorporating frequent feedback and we were interested in whether even a small company like MENDIX has STSCs and if we could detect them.

The system architecture consists of a client system, a workflow server and a modelling server (Fig 5.). The project staff included 15 full-time personnel; 8 full-time developers, 2 project leaders, 2 project managers, 2 sales people and one system administrator. The personnel were divided into 3 teams, with 3 developers, one project leader and one project manager for the client system, 3 developers for the Modelling Server and 3 developers and one project manager for the workflow server. Figure 5 gives a sense of the dependencies as a result of task allocations related to the software architecture of the system. The XML interchange indicates that there exists an input and output dependency between the Server, the XML server and the Client System.

Most of the literature on Socio-Technical dependencies (de Souza et al., 2004; Wagstrom & Herbsleb, 2006) focuses on gathering the dependencies from the recently modified source code (from CVS). We adopted our approach by analysing the source code of the company MENDIX with the help of our tool. We realised that as the company is small most of the dependencies in each section of the architecture (client, xml server, and modeller server) were satisfied by the communication among the developers working on them. Also, knowledge of the technology used in the particular platform (Java, JavaScript or Delphi) was an essential prerequisite for a developer to be working in that part of the project. Due to this fundamental skill requirement we noticed that developers seldom worked on projects or changed code other than their own assigned part of the architecture. As each developer worked on only specific parts of the code, and architecture, there were workflow dependencies between the colleagues due to the architecture. The dependencies due to the XML input and output between the client/server and the servers couldn't be detected by only analysing the call graph and function call dependencies. Thus, we realised that analysing source code of the software product isn't very helpful in analysing the dependencies for a small company like MENDIX. So, we used the technical architecture as a basis to understand the coordination dependencies between the software developers as previously done by Ovaska et. al. (Ovaska et al., 2003).

The data was collected in fall 2006 over a period of 3 months, through participant observation, interviews and gathering work related documents from development tools and communication servers. Among the documents observed were the chat logs, which were stored in XML format. Four weeks of logs of chat transcripts, each week evenly distributed in the 3 month period, were analysed with the help our software tool, TESNA.

All the main developers, project leaders and project managers were interviewed. Among the questions asked in the interview were; who they discuss work related subjects with (advice, discussion and work flow), how much data was exchanged per communication, and, what the mode of communication was. It was ascertained that almost all-technical communication was done through online chat. This was because Mendix uses a dedicated Jabber chat server running for the company (which eliminated wastage of time due to external chats), and developers consider the use of chat more efficient than face-to-face communication. The primary ties in the social networks analysed from the chat log corresponded with those that the interviewees had themselves provided. Further, through participant observation of the software developers (in 6 separate visits lasting a day each) it was ascertained that indeed almost all-technical communication was through the online chat.

The data was analysed and then discussed with the CTO of the company, who doubled as a project manager (roakr in Fig 5.). With the help of our software tool TESNA, the social networks for four different weeks (each with cumulative chat data over the period of a week) of only the developers and project leads/managers were constructed. The chat records were parsed and displayed as social networks by TESNA with the chat ids, as labels for our nodes in the social network. This was also done in compliance with the company policy of protecting the identity of the employees.

We calculated the degree and betweenness centrality (Freeman, 1977) of the nodes and plotted a graph showing its variation over the 3 month period. The resultant diagram was shown to the CTO for his input that was used for a detection of STSCs according to the Betweenness Centrality Match pattern.

TESNA can construct and analyse software metrics from XML logs of chat messages (the chat server being Jabber). Moreover, TESNA displays the different metrics of the social network over a period of time. We used this option to analyse the betweenness centrality of the social networks over the period under study. We took the data back to the CTO once it was displayed and analysed. In this way we could ascertain whether our technique was really useful to the CTO.

The cumulative chat logs were analysed over a period of a week and converted into a social network of the developers and project leaders with the help of our tool (we use JUNG (Madadhain, Fisher, White, & Boey, 2005) to display and calculate metrics). The social network was represented with labels and the number of chat messages exchanged determined the strength of each link. The black links were drawn for the maximum number of chat messages exchanged, and dotted links if the number of chat messages was less than half of the maximum that week.  These links also corresponded to the data we got from the interviews, with respect to who spoke with whom and how much. The system architecture (which didn't change in the period of observation) was then superimposed on the social networks in order to assist the detection of STSCs, according the Conway's Law pattern.

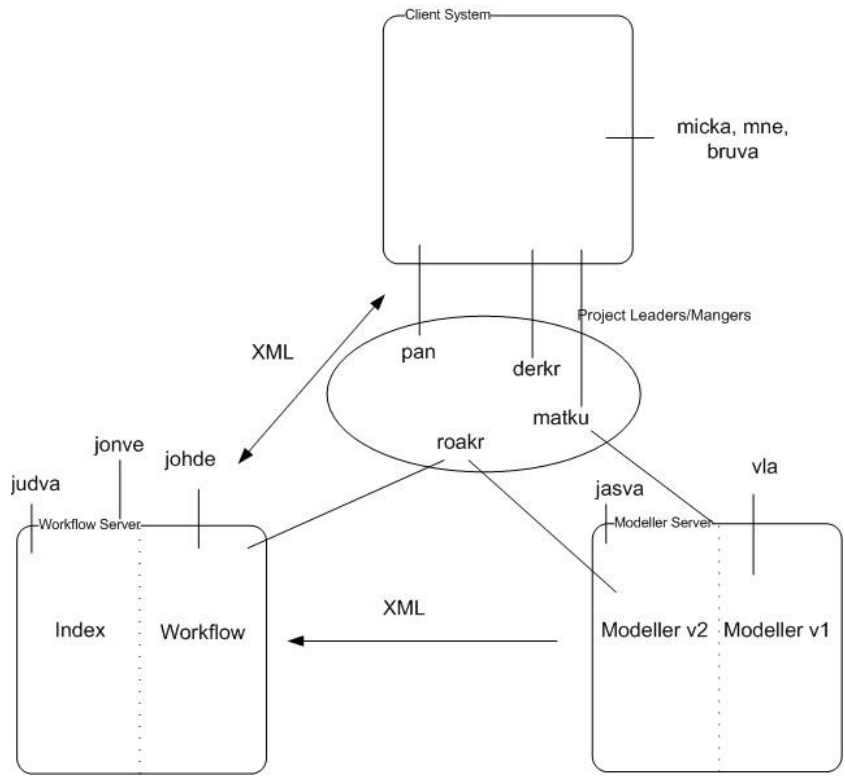

Figure 5: The Software Architecture along with the task responsibilities

## 6. Conway's Law and CTO feedback

The CTO considered Conway's law pattern (Conway, 1968), very important, in his words *"..it is very important that the organization of the company is according to the architecture and I want all the developers to communicate to resolve their problems"*.

So, the CTO was quite pleased when we showed our tool, which maps the social networks to the software architecture of his company's product. When asked how he would expect the social network of the developers and project leads in his company to look, the CTO said

*"I would expect vla, jonve and micka to be central, as they are the Gurus in the work they do and no one knows the functioning of the server, xml server and the client better than them"*

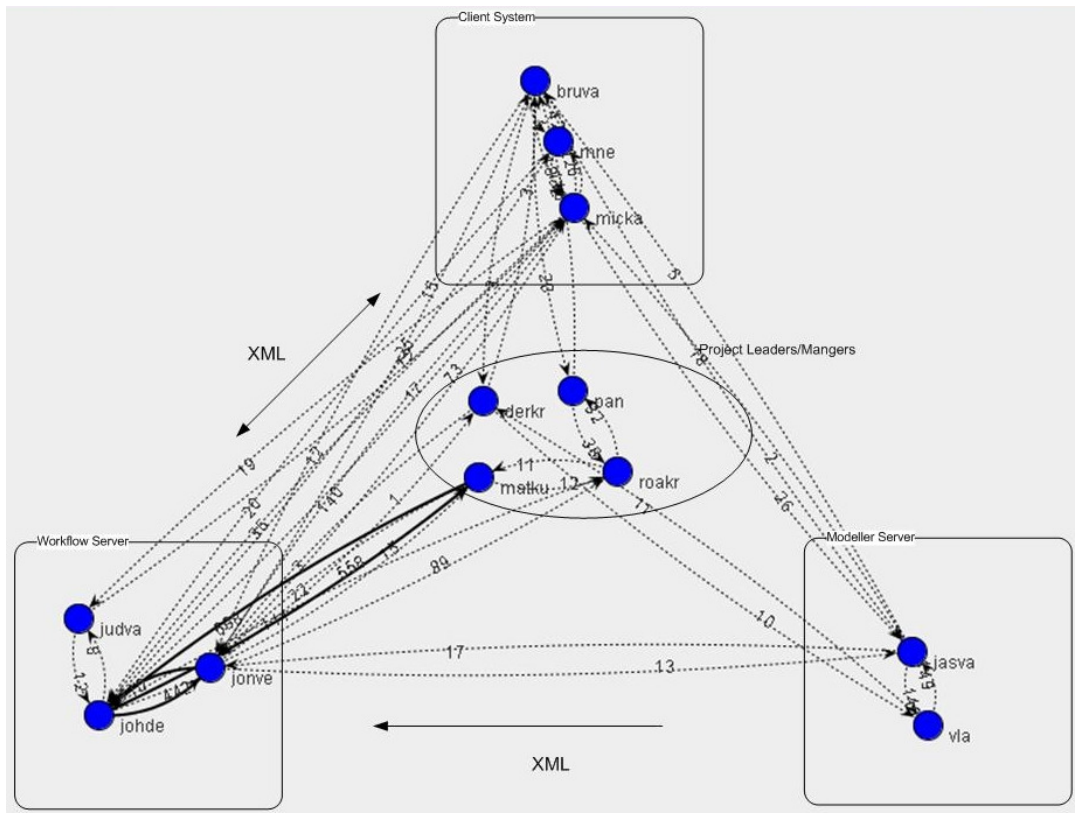

Figure 6: The social network mapped onto the Software Architecture for week I

The social network from week I (Fig 6.) was interesting as the CTO immediately spotted a STSC, which was the missing link between Jonve and Judva, both of whom are developers for the XML server (Fig 5.).

The CTO found the social network from week II (Fig 7) more reasonable than week I, even though there was no connection to johde (who was away doing a project). The three central players were jasva, micka and jonve, which was what he had expected according to the tasks and results in that week. He found that there was little communication with derkr (who is the project manager for the client part of the architecture Fig 5.), which he found odd, as there was some trouble with the client that week.

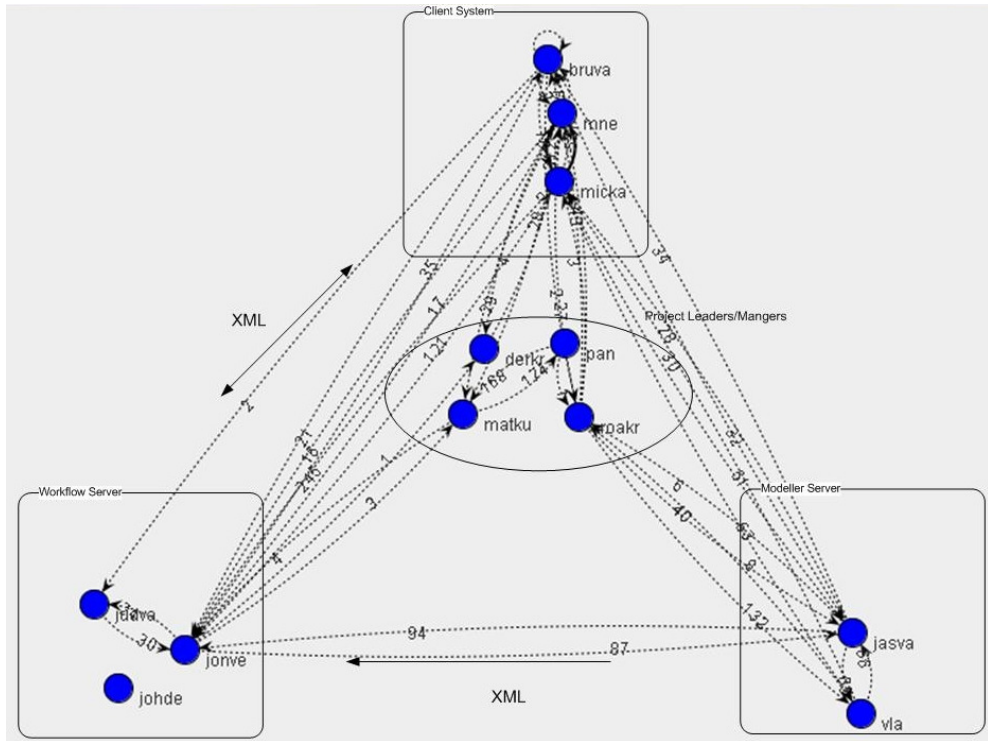

Figure 7: The social Network mapped onto the Software Architecture for week II

Week III (Fig 8.) was interesting as many of the employees were on vacation, and the CTO was interested in how the employees communicated. There was no communication between jasva and micka, as jasva was supposed to work on the Client that week. This could be an indication of a potential problem (or STSC). Also, the CTO found the fact that mne was central quite surprising.

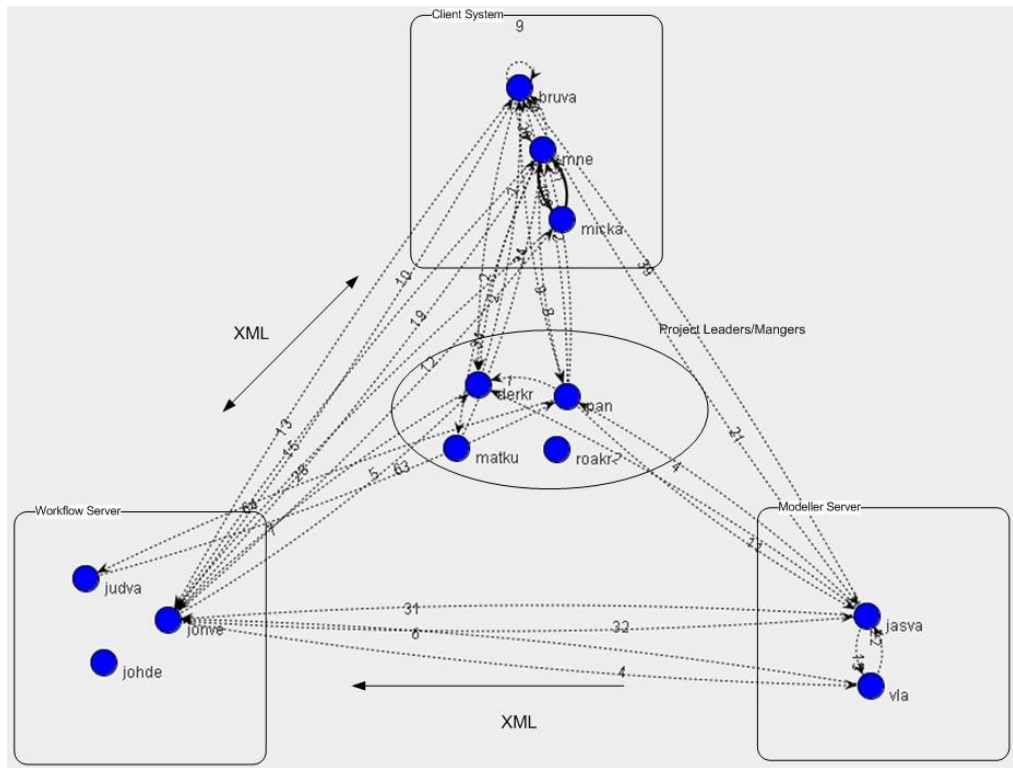

Figure 8: The social Network mapped onto the Software Architecture for week III

In week IV (Fig 9.) the fact that micka was not communicating was surprising as the deadlines were near and it would have been important that he spoke with his fellow client developers. The reason behind pan and matku (having high out-degree) being central was that there was a product shipment on week IV that caused the project leaders to play a more central role. The strong link between jonve and matku was quite odd according to the CTO, as they wouldn't have the need to communicate on technical problems. The fact that bruva had a central role seemed quite odd to the CTO, while the CTO was quite surprised that derkr wasn't communicating much in the week with the shipment deadline.

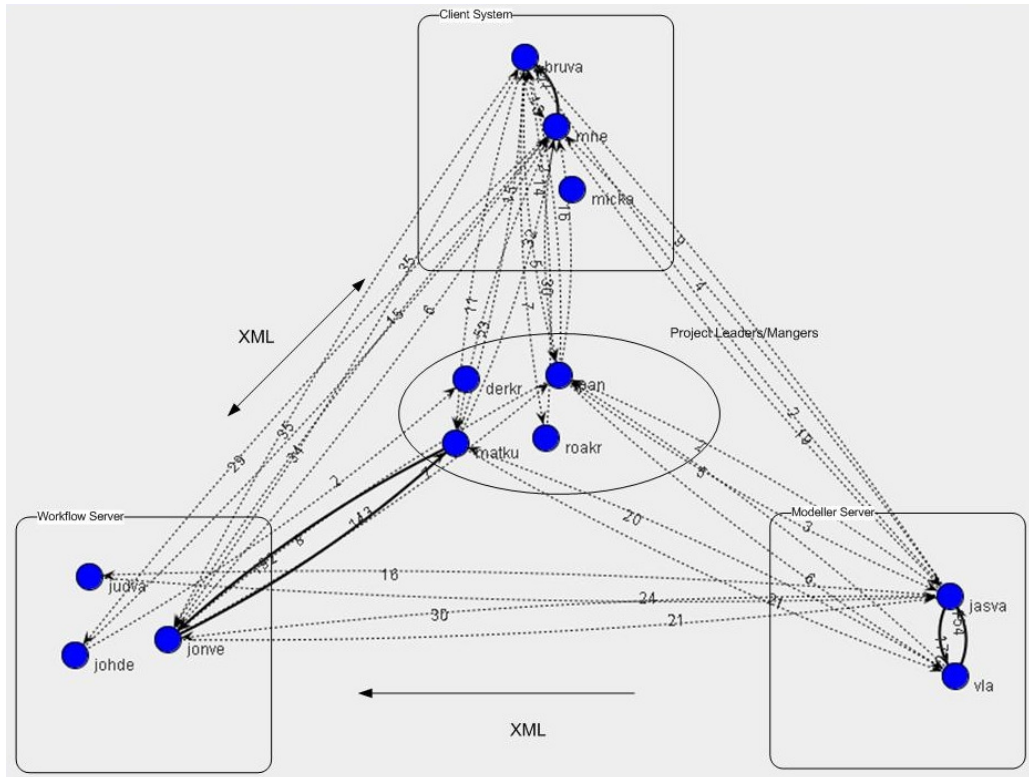

Figure 9: The social Network mapped onto the Software Architecture for week IV

## 7. Betweenness Centrality Match

Centrality index gives us an idea of the potential importance, influence and prominence of an actor in a network. Betweenness refers to the frequency with which a node falls between pairs of other nodes in the network. In other words, betweenness centrality is a measure of, "*the degree that each stands between others, passes messages and thereby gains a sense of importance in contributing to a solution, .. , the greater the betweenness, the greater his or her sense of participation and potency*" (Freeman, 1977). In terms of coordination, betweenness maybe the most appropriate measure of centrality as it provides a measure of the influential control of each node (employee) on the whole networks. This is the reason we used betweenness centrality to analyse potential STSCs.

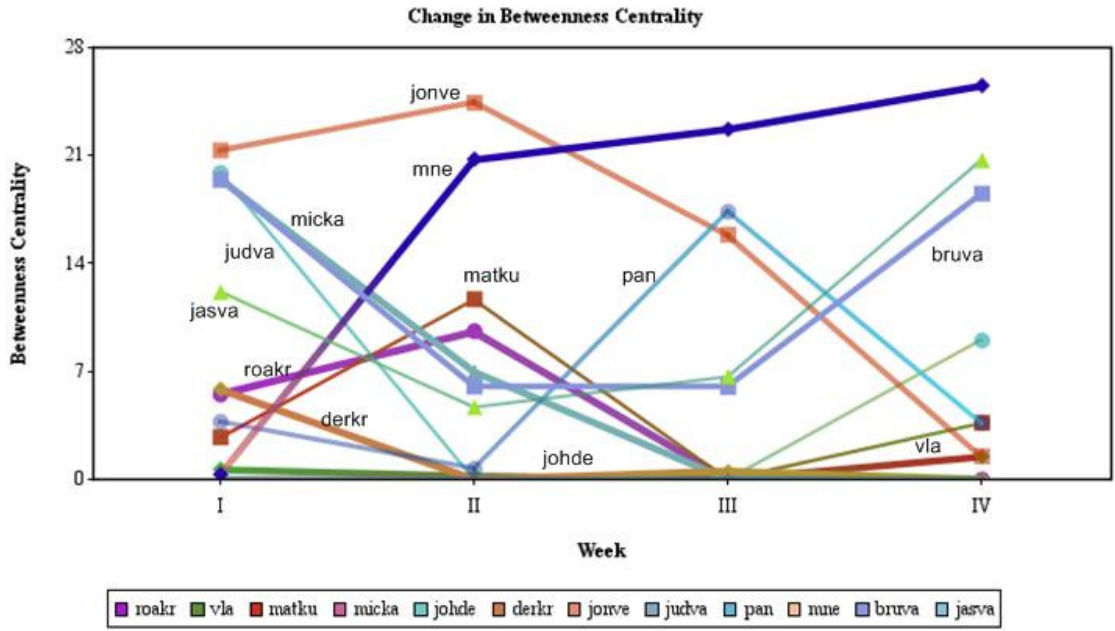

Figure 10: The change in the betweenness centrality over the four weeks

The change in the betweenness centrality index (Freeman, 1977) over the 3 month period can give us an idea of how the most important employee (or the employee who handles most of the communication) in the network changes depending on the tasks at hand. On observing Fig 10, we see that the employees who are important to each part of the software architecture (or the gurus as the CTO called them) namely, jonve, micka and vla were very central during the period around the first week. This period, we realize was exclusively for software development where their expertise was very much (as the CTO named them as the experts in their domain) in demand by fellow developers and project leaders. However, as the project moves towards delivery of the product we find the core developers taking a more passive role in the network while the non-core developers like jasva, bruva and mne as well as the system integration experts take a more central role. This can be explained by the fact that a greater amount of integration and front-end work is required near the delivery deadline.

We also notice that the project leaders and managers (pan and derkr) assume a more central role when the project is nearer to the deadline for delivery to the customer (week IV). This movement to a more central role is required by the project leaders and managers in order to be able to control all the possible contingencies that might crop up near the delivery time. This display of the variation of betweenness centrality index of the social network can also help a manager in recognizing STSCs relevant to different stages in an agile software process. When a person is too central for the wrong reasons, i.e. when a developer is taking responsibility to communicate with the rest of the team, then such a scenario would be a structure clash. For example, the CTO was surprised that mne had a central role in the week III when not much work was required at the client side; he was also surprised that bruva was central in week IV. There is also cause for concern (potential STSC) when two employees working on (or managing) the same part of the architecture (that is being modified) are not communicating with each other, for example micka and derkr were not communicating in any of the weeks under observation.

## 8. Conclusion and Future work

In this paper we have contributed to both the theory and practice of the usage of process patterns in software development. An increasing number of Socio-Technical patterns are becoming available based on experiences and expert opinions. These patterns are potentially useful for managing systems development, but it is difficult and labour intensive for the project manager to select appropriate patterns and keep track of their potential violation. Identifying STSCs can particularly prove difficult when multiple people are responsible for various tasks and when the task requirements keep changing in a dynamic software development environment. We have shown, through our case study, how Socio-Technical

Structure Clashes (STSCs) can be detected during software development. A regular detection of STSCs can help in the management of the software development process.

Though one would expect to find STSCs in large software projects, we were surprised with the presence of STSCs even in a small company like Mendix. This was further reiterated by the fact that the CTO wished to adopt the TESNA method and tool in order to better manage the development process.

We realise that not all software companies would almost exclusively use chat as used in Mendix. In order to overcome this difficulty we plan to keep a track of e-mails (where allowed) as well as meetings and also get feedback on ego-networks with the help of a questionnaire. Apart from such data, we also plan to take interviews in order to corroborate the data, as done in this case study.

In future, we plan to detect technical dependencies at different levels, for example at level of the code (de Souza et al., 2004; Wagstrom & Herbsleb, 2006), those at the level of the architecture and those at the level of work flow. Through the investigation of different dependencies we can gain insights into different possible STSCs. Dependencies due to the code structure we have found are more applicable to large software development organizations.

This paper describes the first steps involved in using this method and tool. We could use other Socio-Technical Patterns (as described in section 3.4) in order to detect different STSCs. In future, we can do a more extensive screening of all the different STSCs that could be detected in the case under inspection. We can also use this technique to validate new and existing Socio-Technical patterns. Future research could also focus on different predictors of STSCs rather than study the outcome of the collaboration to detect STSCs as we have done in this research.

We plan to conduct further case studies to study the presence of STSCs in large software development organizations and inter-organisational development in globally distributed settings. We have also started to investigate open source software development, to see the differences between corporate STSCs and Free/Libre open source STSCs.